\documentclass[a4paper,fleqn]{cas-dc}

\usepackage[authoryear]{natbib}
\usepackage{graphicx}
\usepackage{subfigure} 
\usepackage{multirow} 
\usepackage{threeparttable}
\usepackage{amsmath}
\def\tsc#1{\csdef{#1}{\textsc{\lowercase{#1}}\xspace}}
\tsc{WGM}
\tsc{QE}
\tsc{EP}
\tsc{PMS}
\tsc{BEC}
\tsc{DE}

\begin{document}
\let\WriteBookmarks\relax
\def\floatpagepagefraction{1}
\def\textpagefraction{.001}


\shortauthors{Haoda Li et~al.}

\title [mode = title]{An Improved ESO-Based Line-of-Sight Guidance Law for Path Following of Underactuated Autonomous Underwater Helicopter With Nonlinear Tracking Differentiator and Anti-saturation Controller}                      
\tnotemark[1,2]



%
\author[1]{Haoda Li}[type=editor,
                        auid=000,bioid=1,
                        ]

\fnmark[1]



\credit{Conceptualization of this study, Methodology, Software, Validation, Formal analysis, Writing– original draft, Writing– review editing}

\affiliation[1]{
    organization={Ocean College, Zhejiang University},
    city={Zhoushan},
    postcode={316021}, 
    country={China}}

\author[1]{Zichen Liu}[style=chinese]
\credit{Writing– review editing,Data curation}
\author[1]{Jin Huang}[%
   ]

\credit{Data curation}


\author%
[1]
{Xinyu An}
\credit{Foundation, Writing-review}
\cormark[1]
\ead{anxinyu@zju.edu.cn}
\author%
[1]
{Ying Chen}
\credit{Foundation, Writing-review}


\cortext[cor1]{Corresponding author : Xinyu An}



\begin{abstract}
This paper presents an Improved Extended-state-observer based Line-of-Sight (IELOS) guidance law for path following of underactuated Autonomous Underwater helicopter (AUH) utilizing a nonlinear tracking differentiator and anti-saturation controller. Due to the high mobility of the AUH, the classical reduced-order Extended-State-Observer (ESO) struggles to accurately track the sideslip angle, especially when rapid variation occurs. By incorporating the nonlinear tracking differentiator and anti-saturation controller, the IELOS guidance law can precisely track sideslip angle and mitigate propeller thrust buffet compared to the classical Extended-state-observer based Line-of-Sight (ELOS) guidance law. The performance of ESO is significantly influenced by the bandwidth, with the Improved Extended-State-Observer (IESO) proving effective at low bandwidths where the classical ESO falls short. The paper establishes the input-to-state stability of the closed-loop system. Subsequently, simulation and pool experimental results are showcased to validate the effectiveness of the IELOS guidance law, which outperforms both the Line-of-Sight (LOS) and Adaptive Line-of-Sight (ALOS) guidance laws in terms of performance.
\end{abstract}




\maketitle

\section{Introduction}
In recent years, the motion control of Autonomous Underwater Vehicles (AUVs) has garnered significant interest in realm of ocean technology. Path following plays a crucial role in expanding the application s of AUVs, enabling tasks such as target detection and formation movement. Consequently, an increasing number of researchers are directing their attention towards developing guidance laws for path following, which are essential for converging to the desired path. Among these, the Line-of-Sight (LOS) guidance law has been extensively utilized in path following scenarios due to its simplicity and straightforward implementation.

However the conventional LOS guidance law demonstrates limited performance when AUVs encounter time-varying nonlinear environments including ocean currents and water disturbance, particularly without compensating for the sideslip angle. As a result, Adaptive Line-of-Sight (ALOS) and integrated Line-of-Sight (ILOS) have been introduced to mitigate the impact of non-zero sideslip angles when AUVs track paths.  Both ALOS and ILOS guidance laws have been successfully deployed in various applications, with their superior performance validated by \citep{Fossen2015} and \citep{Caharija2016}. Nevertheless, ALOS and ILOS may experience undesired overshoot or oscillations during the transient phase due to the significant tracking errors. This behavior stems may be caused by their reliance on integral terms, crucial for eliminating steady-state errors, but potentially leading to reduced stability margins due to phase lag.
 
 To tackle these problems, the reduced-order Extended-state-observer based Line-of-Sight (ELOS) guidance law is developed to predict the time-varying sideslip angle, assuming it within a range of no more than 5 degrees \citep{Liu2016}. However the performance of the ELOS guidance law is closely tied to the observer bandwidth, as highlighted in previous studies \citep{Han2009,Gao2003}: higher bandwidth leads to improved performance. Nevertheless, the bandwidth is heavily impacted by sensor accuracy and sampling frequency, and external disturbance, which can often lead to poor performance in underwater scenario. 

Several improved methods based on Line-of-Sight (LOS) guidance laws have been utilized in various scenarios. A prescribed-time Extended State Observer (PTESO) guidance law has been introduced for path tracking of underactuated Unmanned Surface Vehicles (USVs) navigating under time-varying ocean current conditions \citep{Yu2024}. A novel Indirect Adaptive Disturbance Observer (IADO)-based LOS guidance law has been presented for path following of underactuated AUVs in the presence of time-varying ocean currents \citep{Du2023} . R. Rout and  R. Cui introduce a modified LOS guidance law and an adaptive Neural Network (NN) controller for underactuated marine vehicles under uncertainties and constraints\citep{Rout2020}. Y. Xiu presents a finite-time sideslip differentiator based LOS guidance method for robust path following of snake robots \citep{Xiu2023}. S. Zhang proposes two cooperative guidance schemes with impact angle and time constraints for multiple missiles cooperatively intercepting a maneuvering target in the presence or absence of a leader missile \citep{Zhang2021}.
 
The modified LOS guidance law discussed above assume a condition of the low time-varying sideslip angle(usually no more than 5 degrees per second), which may not be suitable to the AUV with high mobility, such as the Autonomous Underwater Helicopter (AUH).
 The AUH is a novel disk-type AUV equipped with features of fixed point hovering, zero-radius turning, and resistance to water flow disturbance\citep{Wang2019,Liu2022,Du2024}. When the AUH follows a curved path, significant sideslip angle occurs (up to 50 degrees per second) with the classical ELOS guidance law in the transient phase. Therefore, utilizing the ELOS guidance law for path following with the AUH is not suitable.

Hence, this paper proposed an Improved Extended-state-observer based Line-of-Sight (IELOS) guidance law for path following of the AUH, especially improving the estimation of the significant time-varying sideslip angle. The main contributions of this paper are outlined as follows:
\begin{enumerate}
    \item To mitigate the outliers of the classical ESO, this paper employs an anti-saturation controller to confine the estimation of sideslip within a reasonable range and eliminate abnormal outliers, thereby ensuring accurate estimation of sideslip angle.
    \item To address the buffeting that may occur in the classical ESO when sideslip angle varies significantly, a nonlinear tracking differentiator is incorporated as a filter to smooth the estimation derived from the reduced-order ESO. This enhancement helps decrease the torque output and buffeting of the propellers.
    \item The IELOS guidance law is introduced to tackle the difficulty of accurately tracking a desired curved path with a highly time-varying sideslip angle. Through validation in both simulation and pool experiments, the proposed guidance law has exhibited superior performance compared to the LOS and ALOS guidance laws. The successful implementation of the IELOS guidance law with a high-mobility AUH highlights its effectiveness, addressing a limitation of the ELOS guidance law.
\end{enumerate}

The structure of this paper is as follows: Section 2 introduces preliminaries, including the coordinate system and tracking-error equation. Section 3 details the design of the IELOS guidance law and includes stability analysis. Section 4 presents the simulation and pool experiment results for validation. Finally, Section 5 offers the conclusion and outlines future work.
\section{Kinematic Preliminaries}
This section presents the coordinate frames, tracking error dynamics, and kinematic differential equations.
\subsection{Coordinate System}
For the conventional torpedo-type AUV and the disk-type AUH, six differnent motion components in the North-East coordinate system are defined as  $\left\{ n \right\} = \left[ {x,y,z,\phi ,\theta ,\psi } \right]$. The body frame is denoted by $\left\{ b \right\} = \left[ {u,v,w_z,p,q,r} \right]$. In practical operations, the pitch angle and roll angle of the AUH can be continuously maintained to zero approximately. Hence, $p = \dot p = \dot q = q = 0$, and the kinematics of the AUH in four-degrees-of-freedom can be simplified to \citep{fossen2011handbook,Li2023}
\begin{equation}
    \left\{ \begin{array}{l}
\dot x = u\cos \psi  - v\sin \psi \\
\dot y = u\sin \psi  + v\cos \psi \\
\dot z = w_z\\
\dot \psi  = r
\end{array} \right.
\label{motion equation}
\end{equation}
where, $w_z$ represents the speed of z-axis.

\begin{figure}
    \centering
    \includegraphics[width=0.8\linewidth]{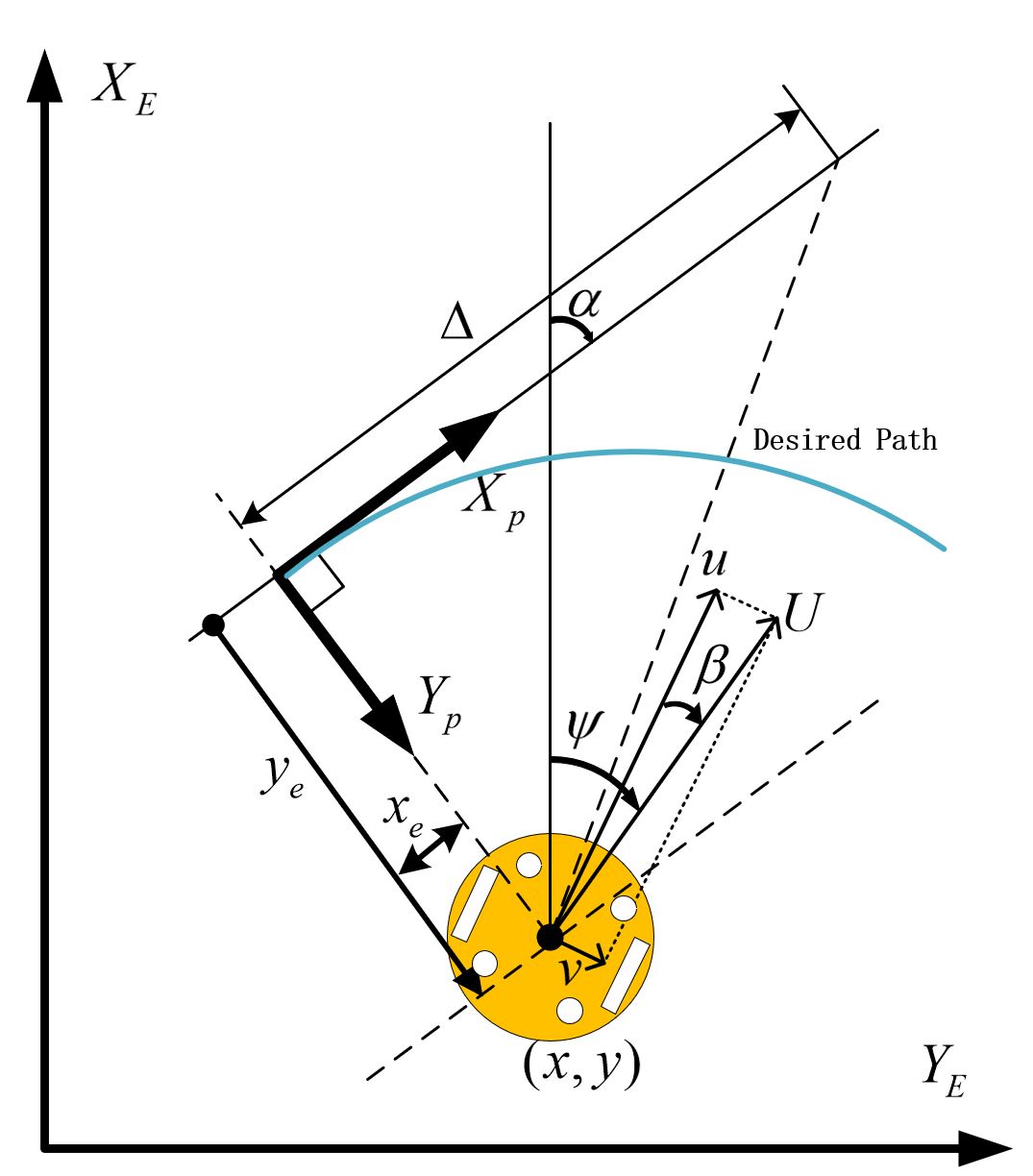}
    \caption{Coordinate system}
    \label{Coordinate system}
\end{figure}

\subsection{Tracking-Error Equations}

The path-tangential coordinate system can be obtained by rotating the North-East coordinate system $\left\{ n \right\}$ and angle $\alpha$ about the downward $z_n$-axis, as shown in Figure \ref{Coordinate system}. The along-track and cross-track errors $({x_e},{y_e})$ expressed in $\left \{ p \right \}$ are given by 

\begin{equation}
\centering
    \left[ {\begin{array}{*{20}{c}}
{{x_e}}\\
{{y_e}}
\end{array}} \right] = \left[ {\begin{array}{*{20}{c}}
{\cos {\alpha _k}}&{\sin {\alpha _k}}\\
{ - \sin {\alpha _k}}&{\cos {\alpha _k}}
\end{array}} \right]\left[ {\begin{array}{*{20}{c}}
{x - {x_k}(w)}\\
{y - {y_k}(w)}
\end{array}} \right]
\end{equation}
where $\left( {{x}_k\left(w\right),{y}_k\left(w\right)} \right)$ denotes the desired geometric path and $w$ represents the path variable. 
The path-tangential angle is given by \({\alpha _k}(w) = \mathop{\rm atan}\nolimits 2({{y'}_k}(w),{x'}_k(w))\), where \({{x'}_k}(w) = \frac{{\partial {x_k}}}{{\partial w}}\), and \({{y'}_k}(w) = \frac{{\partial {y_k}}}{{\partial w}}\).

Then taking the time derivative of $x_e$ and $y_e$
\begin{equation}
    \left\{ \begin{array}{c}
{{\dot x}_e} = \dot x\cos {\alpha _k} + \dot y\sin {\alpha _k} - {{\dot x}_k}(w)\cos {\alpha _k} - {{\dot y}_k}(w)\sin {\alpha _k}\\
 + {{\dot \alpha }_k}[ - (x - {x_k}(w)) \cdot \sin {\alpha _k} + (y - {y_k}(w)) \cdot \cos {\alpha _k}]\\
{{\dot y}_e} =  - \dot x\sin {\alpha _k} + \dot y\cos {\alpha _k} + {{\dot x}_k}(w)\sin {\alpha _k} - {{\dot y}_k}(w)\cos {\alpha _k}\\
 - {{\dot \alpha }_k}[ - (x - {x_k}(w)) \cdot \cos {\alpha _k} + (y - {y_k}(w)) \cdot \sin {\alpha _k}]
\end{array} \right.
\end{equation}


Define $U = \sqrt {{u^2} + {v^2}}  > 0$ as the projected velocity of the AUH on the horizontal plane. $\beta  = {\mathop{\rm atan}\nolimits} 2(v,u)$ represents the sideslip angle,and the $u_p$ is the speed of the virtual reference point expressed by 
\begin{equation}
    {u_p} = \dot w\sqrt {{{x'}_k}^2(w) + {{y'}_k}^2(w)} 
    \label{up}
\end{equation}

Combining
\begin{equation}
\left\{ {\begin{array}{*{20}{c}}
\begin{array}{l}
{{\dot x}_k}(w) = \frac{{\partial {x_k}(w)}}{{\partial w}}\frac{{\partial w}}{{\partial t}} = \dot w \cdot {{x'}_k}(w)\\
 = \dot w\sqrt {{{x'}_k}^2(w) + {{y'}_k}^2(w)}  \cdot \cos {\alpha _k}
\end{array}\\
\begin{array}{l}
{{\dot y}_k}(w) = \frac{{\partial {y_k}(w)}}{{\partial w}}\frac{{\partial w}}{{\partial t}} = \dot w \cdot {{y'}_k}(w)\\
 = \dot w\sqrt {{{x'}_k}^2(w) + {{y'}_k}^2(w)}  \cdot \sin {\alpha _k}
\end{array}
\end{array}} \right.
\end{equation}

Substituting (1), (4) and (5) into (3), the derivatives of the cross-track error and the along-track error can be obtained

\begin{equation}
\centering
    \left\{ \begin{array}{l}
{{\dot x}_e} = U\cos (\psi  - {\alpha _k} + \beta ) + {{\dot \alpha }_k}{y_e} - {u_p}\\
{{\dot y}_e} = U\sin (\psi  - {\alpha _k} + \beta ) - {{\dot \alpha }_k}{x_e}
\end{array} \right.
\label{deriative of cross errors}
\end{equation}

The similar derivation process above can be found in \citep{fossen2017direct,Fossen2023}.
\section{Improved ELOS Guidance Design and analysis}
\subsection{Improved ELOS Guidance}
In this section, an reduced-order ESO with Nonlinear tracking differentiator and anti-saturation controller will be employed to estimate the sideslip angle of the AUH. To simplify the (\ref{deriative of cross errors}), let $\cos \beta  \approx 1$ and $\sin \beta  \approx \beta $. Thus (\ref{deriative of cross errors}) becomes

\begin{figure*}
    \centering
    \includegraphics[width=0.8\linewidth]{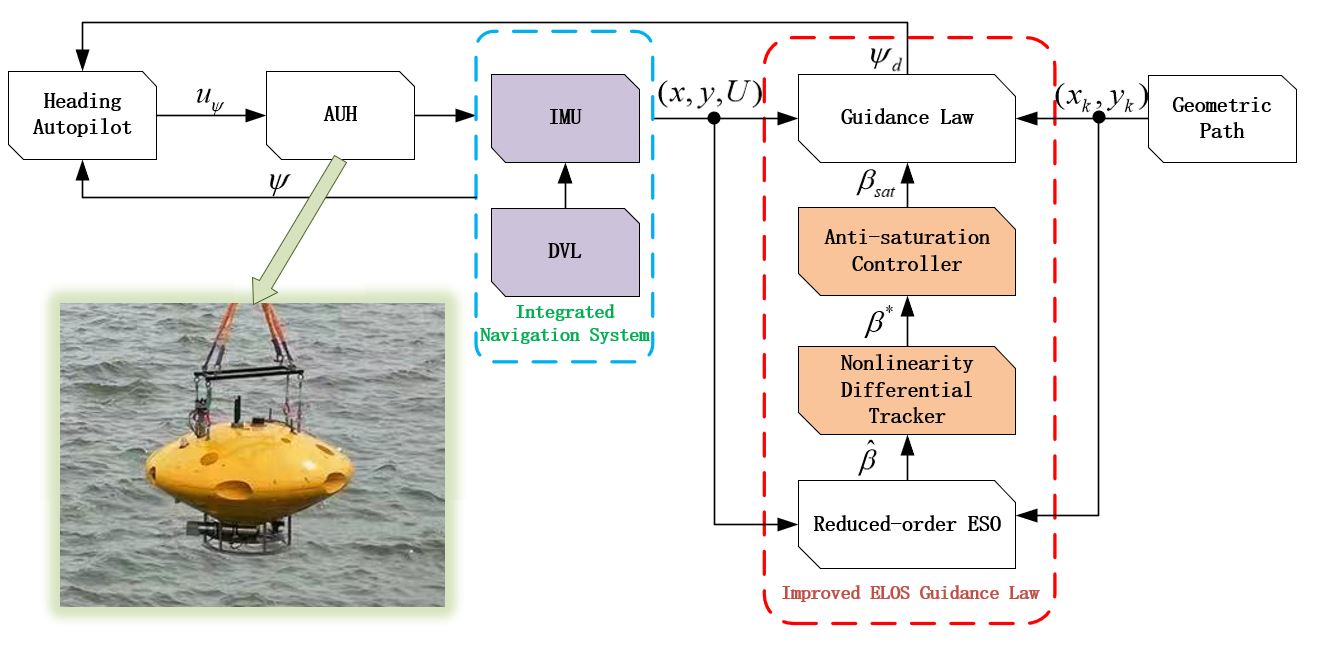}
    \caption{Diagram of the IELOS}
    \label{Controller structure}
\end{figure*}

\begin{equation}
    \left\{ \begin{array}{l}
{{\dot x}_e} = U\cos (\psi  - {\alpha _k}) - U\sin (\psi  - {\alpha _k})\beta  + {{\dot \alpha }_k}{y_e} - {u_p}\\
{{\dot y}_e} = U\sin (\psi  - {\alpha _k}) - U\cos (\psi  - {\alpha _k})\beta  - {{\dot \alpha }_k}{x_e}
\end{array} \right.
\label{simplified deriative of ye}
\end{equation}
Let $g = U\cos (\psi  - {\alpha _k})\beta $.
According \citep{Xingling2015} and \citep{Liu2016}, the reduced-order ESO applied to estimate ${g}$ is presented by 
\begin{equation}
    \left\{ \begin{array}{l}
\dot p =  - kp - {k^2}{y_e} - k\left[ {U\sin ({\psi _d} - {\alpha _k}) - {{\dot \alpha }_k}{x_e}} \right]\\
\hat g = p + k{y_e}
\end{array} \right.
\label{reduced ESO}
\end{equation}
where $\hat g$ is the estimation of $g$. $p$ denotes the auxiliary state of the observer, $k$ represents the observer gain. $\psi_d$ denotes the desired heading decided by guidance law. Hence, the estimation of $\beta$ can be deprived as $\hat \beta$.

 In practice, the estimation $\hat \beta$ obtained by reduced-order ESO often leads to the buffeting effect, which can have adverse effects on the autopilot controller and propeller actuators. To mitigate this issue, a tracking differentiator is utilized as a filter to reduce high-frequency oscillations and smooth the torque output of the propellers. Initially proposed by \citep{Han2009}, a noise-tolerant time optimal
control (TOC)-based tracking differentiator allows one
to avoid a setpoint jump in the emerging active disturbance
rejection controller. The advantage of this tracking differentiator is to maintain a greater level of smoothness than
 sliding mode-based differentiators \citep{Zhang2019}. The tracking differentiator is presented as follows: 
\begin{equation}
\left\{ \begin{array}{l}
{{\dot v}_1} = {v_2}\\
{{\dot v}_2} = {\mathop{\rm fhan}\nolimits} \left( {{v_1},{v_2},r,h} \right)
\end{array} \right.
\end{equation}

Define ${\mathop{\rm fsg}\nolimits} \left( {x,d} \right) = {{\left( {{\mathop{\rm sign}\nolimits} \left( {x + d} \right) - {\mathop{\rm sign}\nolimits} \left( {x - d} \right)} \right)} \mathord{\left/
 {\vphantom {{\left( {{\mathop{\rm sign}\nolimits} \left( {x + d} \right) - {\mathop{\rm sign}\nolimits} \left( {x - d} \right)} \right)} 2}} \right.
 \kern-\nulldelimiterspace} 2}$, ${\mathop{\rm fhan}\nolimits} \left( {{v_1},{v_2},r,h} \right)$ can be expressed as
 \begin{equation}
{\mathop{\rm fhan}\nolimits} \left( {{v_1},{v_2},r,h} \right) = \left\{ \begin{array}{l}
d = r{h^2}\\
{a_0} = h{v_2}\\
y = {v_1} + {a_0}\\
{a_1} = \sqrt {d\left( {d + 8\left| y \right|} \right)} \\
{a_2} = {{{a_0} + {\mathop{\rm sign}\nolimits} \left( y \right)\left( {{a_1} - d} \right)} \mathord{\left/
 {\vphantom {{{a_0} + {\mathop{\rm sign}\nolimits} \left( y \right)\left( {{a_1} - d} \right)} 2}} \right.
 \kern-\nulldelimiterspace} 2}\\
a = \left( {{a_0} + y} \right){\mathop{\rm fsg}\nolimits} \left( {y,d} \right) + {a_2}\left( {1 - {\mathop{\rm fsg}\nolimits} \left( {y,d} \right)} \right)\\
{\mathop{\rm fhan}\nolimits}  =  - r\left( {{a \mathord{\left/
 {\vphantom {a b}} \right.
 \kern-\nulldelimiterspace} b}} \right){\mathop{\rm fsg}\nolimits} \left( {a,d} \right)\\
\begin{array}{*{20}{c}}
{}&{ - r{\mathop{\rm sign}\nolimits} \left( a \right)\left( {1 - {\mathop{\rm fsg}\nolimits} \left( {a,d} \right)} \right)}
\end{array}
\end{array} \right.
\label{TD}
\end{equation}

For second-order tracking differentiator, all solutions of the above formula (in Filippov sense) are bounded and satisfied
\begin{equation}
   \left\{ \begin{array}{l}
\mathop {\lim }\limits_{t \to \infty } {v_1}(t) = 0\\
\mathop {\lim }\limits_{t \to \infty } {v_2}(t) = 0
\end{array} \right.
\label{limits}
\end{equation}
where,${v_1} = {\beta ^*}\left( t \right) - \hat \beta \left( t \right)$, ${\beta ^*}\left( t \right)$ is the tracking value of the $\hat \beta$. Hence, equation (\ref{limits}) can be deprived to $\mathop {\lim }\limits_{t \to \infty } {\beta ^*}\left( t \right) = \hat \beta \left( t \right)$.

In equation(\ref{TD}), $r>0$ represents the speed factor of the tracking diferentiator, with a larger $r$ indicating that the value $\beta ^{*}$ can track the $\hat \beta$ more rapidly. The parameter $h$ serves as a filtering factor for the tracking diferentiator, when the input signal is contaminated with noise, an appropriate $h$ ensures a clean, accurate signal.

The convergence time and effectiveness of the observer in the reduced-order ESO are significantly impacted by the bandwidth parameter $k$, where lower bandwidth $k$ results in poor observer performance. However, environmental disturbance, sensor sampling frequency, and accuracy constraints often limit the bandwidth value $k$ maintaining at a low level in practical operations, Consequently, the estimation $\hat \beta$ may not accurately track the actual value of $\beta$, and in severe cases, the estimate may exceed the actual value. Given that the sideslip angle $\beta$ should fall within the range of $-\pi$ to $\pi$, the anti-saturation controller introduced here aims to confine the estimated value $\hat \beta$ avoiding system instability due to abnormal outliers. The anti-saturation controller is expressed as \citep{Chen2022}:
\begin{equation}
{\beta _{sat}} = {\mathop{\rm sat}\nolimits} ({\beta ^*}) = \left\{ {\begin{array}{*{20}{c}}
{{\beta _{\mathop{\rm m}\nolimits} }{\mathop{\rm sign}\nolimits} \left( {{\beta ^*}} \right)}&{\left| {{\beta ^*}} \right| \ge {\beta _{\mathop{\rm m}\nolimits} }}\\
{{\beta ^*}}&{\left| {{\beta ^*}} \right| < {\beta _m}}
\end{array}} \right.
\end{equation}
where ${\mathop{\rm sat}\nolimits} \left(  \cdot  \right)$ and ${\mathop{\rm sign}\nolimits} \left(  \cdot  \right)$ represent the saturation and unit sign function, respectively. ${\beta _m}$ denotes the known magnitude of the saturation limit and ${\beta _m} = \pi$. 
\begin{figure*}[!tbh]
    \centering
    \includegraphics[width=1\linewidth]{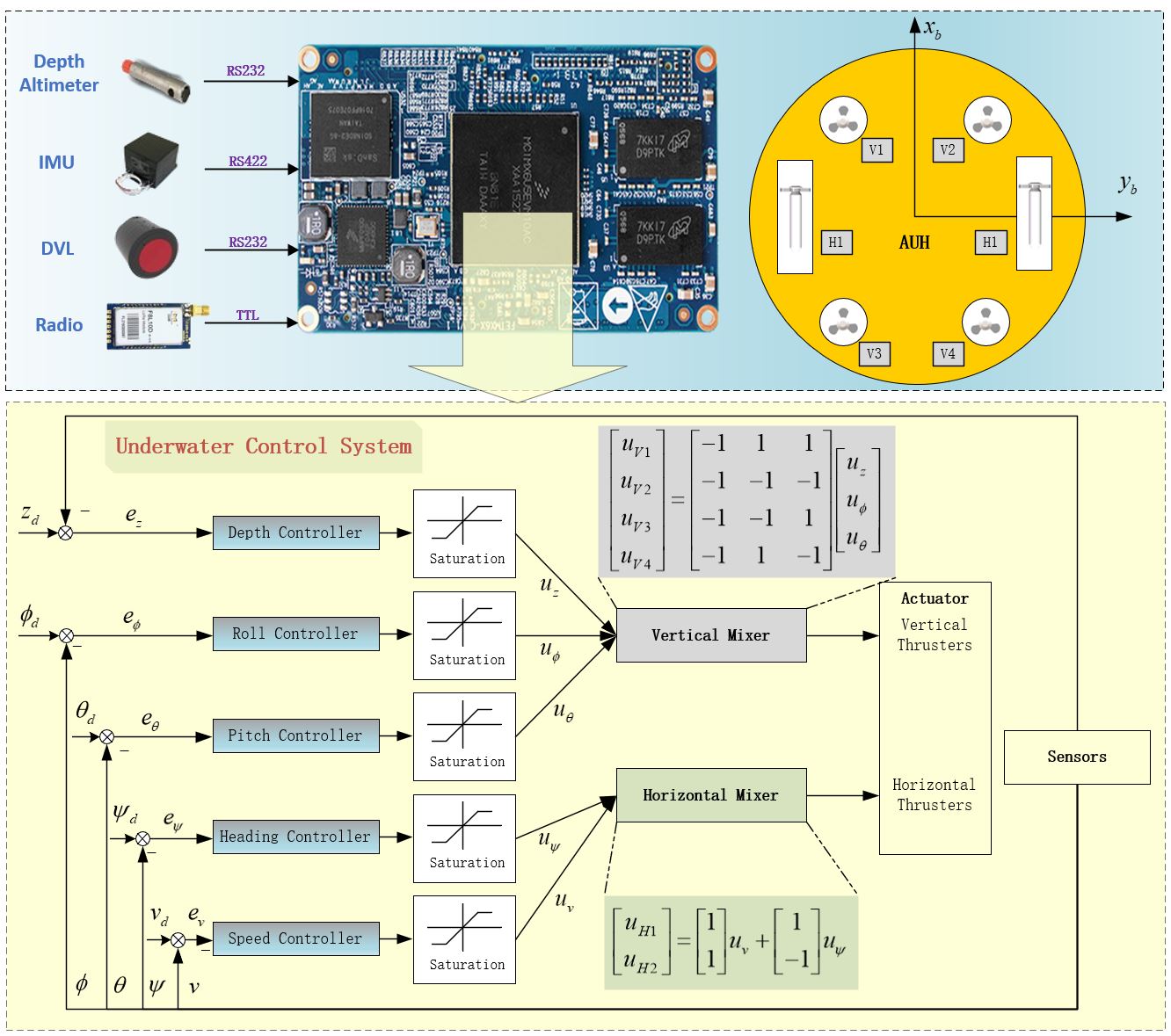}
    \caption{Controller system of AUH}
    \label{Controller system of AUH}
\end{figure*}
\subsection{Guidance law design}
The IELOS guidance law presented in this paper incorporates a nonlinear tracking differentiator and anti-saturation within the ESO-based LOS guidance law. The diagram of the IELOS guidance law and the photograph of the AUH are depicted in Figure\ref{Controller structure}. Subsequently, the stability analysis will be presented. 
To start with, the fundamental guidance law is proposed as
\begin{equation}
    {\psi _d} = {\alpha _k}(w) + \arctan (\frac{{ - {y_e}}}{\Delta } - \beta_{sat} )
\end{equation}

Define
\begin{equation}
    {u_p} = U\cos ({\psi _d} - {\alpha _k}) + \kappa {x_e}
    \label{up_xe}
\end{equation}

Combing (\ref{up}), $\dot w$ can be deprived as
\begin{equation}
    \dot w = \frac{{U\cos ({\psi _d} - {\alpha _k}) + \kappa {x_e}}}{{\sqrt {{{x'}_k}^2(w) + {{y'}_k}^2(w)} }}
    \label{w derivation}
\end{equation}

Substituting (\ref{up_xe}) into the first equality of (\ref{simplified deriative of ye}) gives
\begin{equation}
    {\dot x_e} =  - \kappa {x_e} - U\cos (\psi  - {\alpha _k})\beta  + {\dot \alpha _k}{y_e}
\end{equation}

Known triangle inequality
\begin{equation}
\left\{ {\begin{array}{*{20}{c}}
{\sin ({{\tan }^{ - 1}}( - \frac{{{y_e}}}{\Delta } - {\beta _{sat}})) =  - \frac{{{y_e} + \Delta {\beta _{sat}}}}{{\sqrt {{\Delta ^2} + {{({y_e} + \Delta {\beta _{sat}})}^2}} }}}\\
{\cos ({{\tan }^{ - 1}}( - \frac{{{y_e}}}{\Delta } - {\beta _{sat}})) = \frac{\Delta }{{\sqrt {{\Delta ^2} + {{({y_e} + \Delta {\beta _{sat}})}^2}} }}}
\end{array}} \right.
\end{equation}

The time derivative of $y_e$ is deprived to
\begin{equation}
\begin{array}{c}
{{\dot y}_e} =  - U\frac{{({y_e} + \Delta {\beta _{sat}})}}{{\sqrt {{\Delta ^2} + {{({y_e} + \Delta {\beta _{sat}})}^2}} }} + U\frac{{\Delta {\beta _{sat}}}}{{\sqrt {{\Delta ^2} + {{({y_e} + \Delta {\beta _{sat}})}^2}} }} - {{\dot \alpha }_k}{x_e}\\
 =  - \frac{{U{y_e}}}{{\sqrt {{\Delta ^2} + {{({y_e} + \Delta {\beta _{sat}})}^2}} }} - \frac{{U\Delta \left( {{\beta _{sat}} - \beta } \right)}}{{\sqrt {{\Delta ^2} + {{({y_e} + \Delta {\beta _{sat}})}^2}} }} - {{\dot \alpha }_k}{x_e}
\end{array}
\end{equation}

Finally, the error dynamics in terms of $x_e$ and $y_e$ become


\begin{equation}
\left\{ \begin{array}{l}
{{\dot x}_e} =  - \kappa {x_e} - U\cos (\psi  - {\alpha _k})\beta  + {{\dot \alpha }_k}{y_e}\\
{{\dot y}_e} =  - \frac{{U{y_e}}}{{\sqrt {{\Delta ^2} + {{({y_e} + \Delta {\beta _{sat}})}^2}} }} - \frac{{U\Delta \left( {{\beta _{sat}} - \beta } \right)}}{{\sqrt {{\Delta ^2} + {{({y_e} + \Delta {\beta _{sat}})}^2}} }} - {{\dot \alpha }_k}{x_e}
\label{ye_beta_sat}
\end{array} \right.
\end{equation}

In this section, we introduce the concept of input-state stability. For a general nonlinear system, the property of being ISS is typically established by demonstrating the existence of an ISS-Lyapunov function \citep{Sontag1995,drazin1992nonlinear}.

\textbf{Lemma 1}: Consider the following general nonlinear system $\dot x = f\left( {x,t,u} \right)$. A smooth function $V:\mathbb{R}{^n} \to \mathbb{R}{_{ \ge 0}}$ is called an ISS-Lyapunov function for system above. if there exist ${\mathcal{{\cal K}}_\infty }$-function $\alpha_1$, $\alpha_2$ and ${\mathcal{{\cal K}} }$-function $\alpha_3$ and $\chi$, such that ${\alpha _1}\left( {\left\| x \right\|} \right) \le V\left( {t,x} \right) \le \left( {\left\| x \right\|} \right)$ and ${{\partial V} \mathord{\left/
 {\vphantom {{\partial V} {\partial t}}} \right.
 \kern-\nulldelimiterspace} {\partial t}} + \left( {{{\partial V} \mathord{\left/
 {\vphantom {{\partial V} {\partial x}}} \right.
 \kern-\nulldelimiterspace} {\partial x}}} \right) \cdot f\left( {t,x,u} \right) \le  - {\alpha _3}\left( {\left\| x \right\|} \right),\forall \left\| x \right\| \ge \chi \left( {\left\| u \right\|} \right)$ ,where $x \in \mathbb{R} {^n},u \in \mathbb{R} {^m}$. Then, the nonlinear system above is ISS with the input being $u$, state being $x$, and $\gamma  = {\alpha _1}^{ - 1} \circ {\alpha _2} \circ \chi $.
 
 To move on, the following assumptions are required.

\textbf{Assumption 1}: The heading autopilot tracks the guidance heading angle perfectly such that $\psi  = {\psi _d}$.

\textbf{Assumption 2}: The tracking differentiator follows the estimation $ \hat \beta$ obtained by reduced-order ESO perfectly such that $\beta{^*} = \hat \beta$.

Combing the assumption above, the following lemma presents the stability of (\ref{ye_beta_sat}).

\textbf{Lemma 2}:  system(\ref{ye_beta_sat}) with the states being $x_e$ and $y_e$, the inputs being $\beta$, is ISS, provided that
\begin{equation}
    \begin{array}{*{20}{c}}
{\kappa  - \frac{U}{{2{\varepsilon _1}}} > 0,}&{c - \frac{U}{{2{\varepsilon _2}}} > 0}
\end{array}
\end{equation}
\textbf{Proof}: Consider Lyapunov function candidate

\begin{align}
\centering
    V = \frac{1}{2}{x_e}^2 + \frac{1}{2}{y_e}^2
\end{align}

The time derivative of $V$ is 
\begin{equation}
\centering
\begin{array}{c}
\dot V =  - \kappa {x_e}^2 - U\sin ({\psi _d} - {\alpha _k})\beta {x_e}\\
 - \frac{{U{y_e}^2}}{{\sqrt {{\Delta ^2} + {{({y_e} + \Delta {\beta _{sat}})}^2}} }} - \frac{{U{y_e}\Delta \left( {{\beta _{sat}} - \beta } \right)}}{{\sqrt {{\Delta ^2} + {{({y_e} + \Delta {\beta _{sat}})}^2}} }}
\end{array}
\end{equation}

Next, we define
\begin{equation}
\centering
    c = \frac{U}{{\sqrt {{\Delta ^2} + {{({y_e} + \Delta  \beta_{sat} )}^2}} }}    
\end{equation}

Using the known inequations
\begin{equation}
\centering
\left\{ \begin{array}{c}
 - U\cos ({\psi _d} - {\alpha _k})\beta {x_e} \le \frac{U}{{2{\varepsilon _1}}}{\left| {{x_e}} \right|^2} + \frac{{U{\varepsilon _1}}}{2}{\left| \beta  \right|^2}\\
 - {y_e}\frac{{U\Delta \left( {{\beta _{sat}} - \beta } \right)}}{c} \le \frac{1}{{2{\varepsilon _2}}}{\left| {{y_e}} \right|^2} + \frac{{{\varepsilon _2}}}{2}{\left| {\frac{{U\Delta \left( {{\beta _{sat}} - \beta } \right)}}{c}} \right|^2}
\end{array} \right.
\end{equation}

The time derivative of $V$ is deprived to
\begin{equation}
\centering
\begin{array}{c}
\dot V \le  - (\kappa  - \frac{U}{{2{\varepsilon _1}}}){x_e}^2 - (c - \frac{1}{{2{\varepsilon _2}}}){y_e}^2 + \frac{{U{\varepsilon _1}}}{2}{\left| \beta  \right|^2}\\
 + \frac{{{\varepsilon _2}}}{2}{\left| {\frac{{U\Delta \left( {{\beta _{sat}} - \beta } \right)}}{c}} \right|^2}
\end{array}
\end{equation}

Noting that
\begin{equation}
\centering
    h = \min \left\{ {(\kappa  - \frac{U}{{2{\varepsilon _1}}}),(c - \frac{1}{{2{\varepsilon _2}}})} \right\} > 0
\end{equation}

Then
\begin{align}
\centering
 \begin{array}{l}
\dot V \le  - h{\left\| E \right\|^2} + \frac{{U{\varepsilon _1}}}{2}{\left| \beta  \right|^2} + \frac{{{\varepsilon _2}}}{2}{\left| {\frac{{U\Delta \left( {{\beta _{sat}} - \beta } \right)}}{c}} \right|^2}\\
 \le  - \frac{h}{2}{\left\| E \right\|^2} - \left( {\frac{h}{2}{{\left\| E \right\|}^2} - \frac{{U{\varepsilon _1}}}{2}{{\left| \beta  \right|}^2} - \frac{{{\varepsilon _2}}}{2}{{\left| {\frac{{U\Delta \left( {{\beta _{sat}} - \beta } \right)}}{c}} \right|}^2}} \right)
\end{array}
\end{align}
where $E = {\left[ {{x_e},{y_e}} \right]^T}$.

Combining the definitions of $\beta$ and $\beta_{sat}$, $\left| {{\beta _{sat}} - \beta } \right| \le 2\pi $. Thus

\begin{align}
\begin{array}{*{20}{c}}
{\dot V \le  - \frac{h}{2}{{\left\| E \right\|}^2},}&{\forall E \ge \sqrt {\frac{{U{\varepsilon _1}}}{h}} \left| \beta  \right| + }
\end{array}\sqrt {\frac{{{\varepsilon _2}}}{h}} \frac{{U\Delta 2\pi }}{c}
\end{align}
 it follows that the error system (\ref{ye_beta_sat}) is ISS.
 
 Choose ${\alpha _{c1}}\left( s \right) = {\alpha _{c2}}\left( s \right) = \left( {{1 \mathord{\left/
 {\vphantom {1 2}} \right.
 \kern-\nulldelimiterspace} 2}} \right){s^2}$,
  then there exist class $\mathcal{{\cal K}}L$ function $\sigma_c$ and class $\mathcal{{\cal K}}$ function $\gamma _{c1}$, such that
  \begin{equation}
      \left\| {e\left( t \right)} \right\| \le {\sigma _c}\left( {\left\| {e\left( 0 \right)} \right\|,t} \right) + {\gamma _{c1}}\left( {\left| \beta  \right|} \right)
  \end{equation}


\section{Case Study With Autonomous Underwater Helicopter}
This section introduces the basic information of the AUH, and presents the path following results of simulation and pool experiments using LOS, ALOS, ELOS, and IELOS guidance laws.

\subsection{Introduction of Autonomous Underwater Helicopter}
 In contrast to the conventional torpedo-shaped AUVs, the AUH using a novel disk-shaped design that highlight its exceptional mobility capabilities. The specific technical details and specifications of the AUH are comprehensively outlined in Table \ref{Specification}, while its intricate controller system is visually presented in Figure \ref{Controller system of AUH}. The device is equipped with a total of six propellers, with two horizontal propellers assigned for straight and steering functions, and four vertical propellers designated for depth, roll angle, and pitch angle control. Consequently, the AUH can be classified as a typical underactuated system.
 \begin{table}[width=.9\linewidth,cols=3,pos=h]
\caption{Specification of the Autonomous Underwater Helicopter}
\begin{tabular*}{\tblwidth}{@{} LLL@{} }
    \toprule
          Parameters   &  value & units \\
    \midrule
         Diameter &  1.8 & m\\
         Height & 1.6 & m\\
         Weight & 750 & kg\\
         Max Thrust & 400 & N\\
         Max Depth & 1500 & m\\
   \bottomrule
\end{tabular*}
\label{Specification}
\end{table}

\begin{table}[width=1\linewidth,cols=4,pos=h!]
\begin{threeparttable}
\caption{The parameters of six cases in simulation.}\label{tbl1}
\begin{tabular*}{\tblwidth}{@{} L|L|L|L|L|L@{} }
\toprule
LOS & ALOS & \multicolumn{2}{c|}{ELOS}& \multicolumn{2}{c}{IELOS} \\
\midrule
$case1$ & $case2$ &$case3$ & $case4$ & $case5$ &$case6$\\
$\Delta = 2$ & $\Delta = 2$ &$\Delta = 2$  &$\Delta = 2$&$\Delta = 2$& $\Delta = 2$ \\
$\kappa =10$& $\kappa =10$ &$\kappa =10$ & $\kappa^{1} =0.3$ &$\kappa =10$ &$\kappa =10$ \\
 & $l=0.005$ & $k=10$ & $k=2$ & $k=10$ &$k=2$\\
 &  &  &  & $r=30$ &$r=30$\\
 &  & &  & $h=0.05$ &$h=0.05$\\
\bottomrule
\end{tabular*}
\begin{tablenotes}
\item[1] $\kappa$ have to be set to 0.3 since the ELOS is incompatible with $k=2$ and $\kappa =10$.
\end{tablenotes}
\end{threeparttable}
\end{table}
\subsection{Simulation of a circular path}

\begin{figure}
    \centering
    \includegraphics[width=1\linewidth]{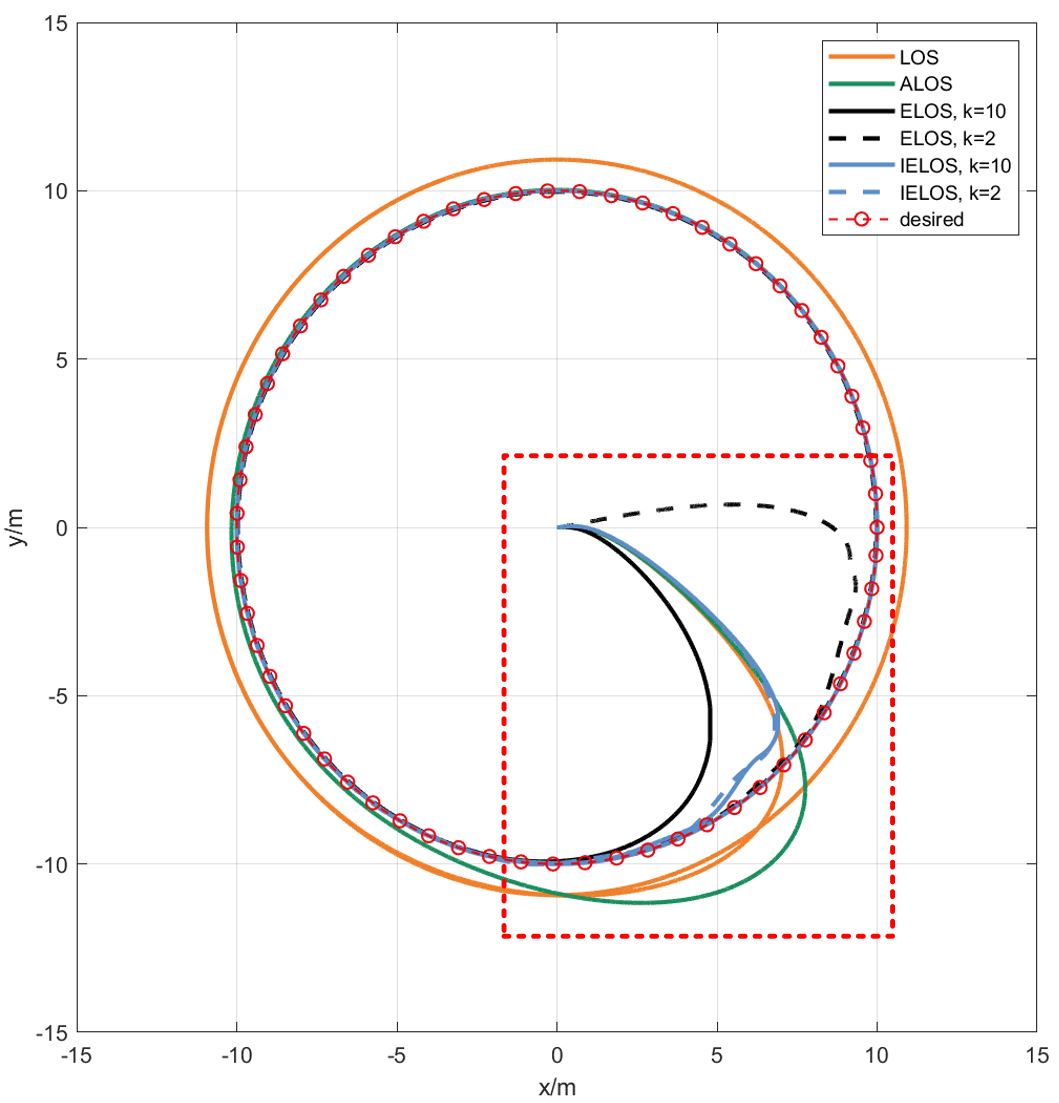}
    \caption{The trajectory of various guidance laws in simulation}
    \label{simulation traj}
\end{figure}
\begin{figure*}
    \centering
\includegraphics[width=0.8\linewidth]{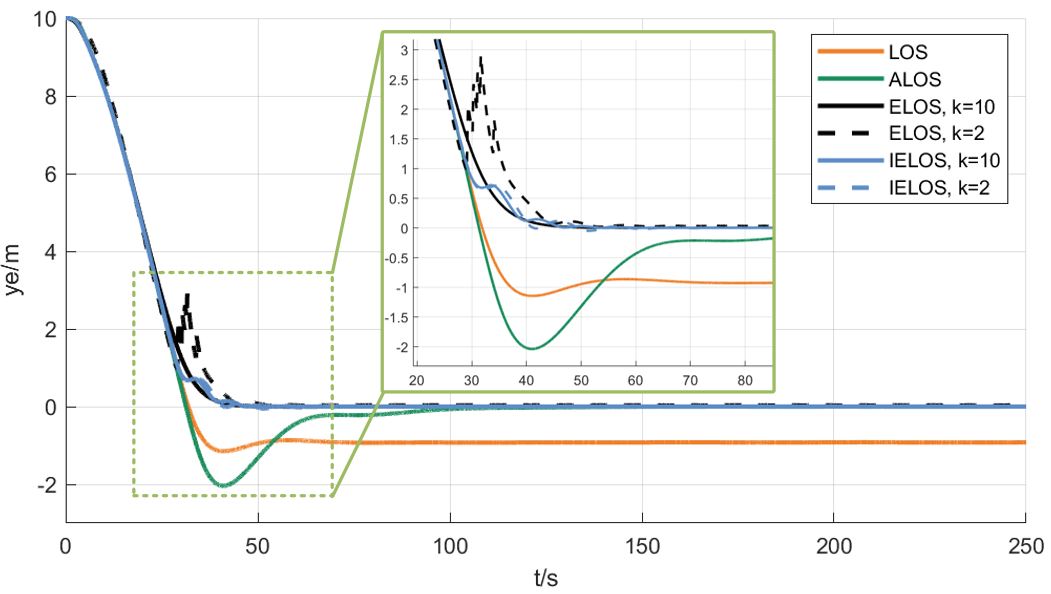}
    \caption{The $y_e$ of various guidance laws in simulation}
    \label{simulation ye}
\end{figure*}

In this paper, the comparative simulation results of LOS, ALOS, ELOS and IELOS are presented as follows. To begin with, the LOS guidance law is expressed as
\begin{equation}
    {\psi _d} = {\alpha _k}(w) + \arctan (\frac{{ - {y_e}}}{\Delta })
\end{equation}

As a contrast, the ALOS guidance law proposed in \citep{Fossen2015} is as follows.
\begin{equation}
    \left\{ \begin{array}{l}
{\psi _d} = {\alpha _k}(w) + \arctan \left( {\frac{{ - {y_e}}}{\Delta } - \hat \beta } \right)\\
\begin{array}{*{20}{c}}
{\dot {\hat \beta}  = l\frac{{{y_e}U\Delta }}{{\sqrt {{\Delta ^2} + {{\left( {{y_e} + \Delta \hat \beta } \right)}^2}} }}}&{l > 0}
\end{array}
\end{array} \right.
\end{equation}

Obviously, the parameter $\Delta$ and $k$ are both employed in LOS, ALOS, ELOS, and IELOS guidance laws. It is common practice to set the corresponding parameters of different guidance laws to the same value in order to facilitate a more straightforward comparison. Therefore, the parameters $\Delta$ and $\kappa$ are all typically set to 2 and 10, respectively, unless the scenario involves $k=2$ in ELOS case ($case4$). In instances where $k=2$ and $\kappa=10$ in ELOS guidance law, significant variations of $\hat \beta$ are observed, leading to a considerable deviation from the true value and causing the simulation code to crash. Consequently, when bandwidth $k$ is equal to 2 in ELOS guidance law, the $\kappa$ have to be adjusted to 0.3. In order to clarify the description process, the six cases in the simulation are named as $case1$ $\sim $ $case6$, respectively. The parameters of various cases are listed in Table \ref{parameters in experiment}.

In these simulation cases, a circular trajectory with a radius of 10 m is designed as the desired path, which is expressed as
\begin{equation}
    \left\{ \begin{array}{l}
x_k\left( w \right) = R\cos \left( w \right)\\
y_k\left( w \right) = R\sin \left( w \right)
\end{array} \right.
\end{equation}
where $R$ denotes the radius of the circular path.
\begin{table}[width=.9\linewidth,cols=3,pos=h!]
\caption{The MAE of $x_e$ and $y_e$ in simulation}
\begin{tabular*}{\tblwidth}{@{} LLL@{} }
    \toprule
             &  $x_e$ (unit: m) & $y_e$ (unit: m)\\
    \midrule
         LOS (case 1) &  $0.0310$ & $1.4261$\\
         ALOS (case 2)& $0.0192$ & $0.9265$\\
         ELOS, k=10 (case 3)& $0.0143$ & $0.7753$\\
         ELOS, k=2 (case 4)& $0.3809$ & $0.8365$\\
         IELOS, k=10 (case 5)& $0.0143$ & $0.7656$\\
         IELOS, k=2 (case 6)& $0.0139$ & $0.7682$ \\
   \bottomrule
\end{tabular*}
\label{MAE in simulation}
\end{table}
\begin{figure}[pos=h!]
\centering
\subfigure[]{
\includegraphics[width=3in]{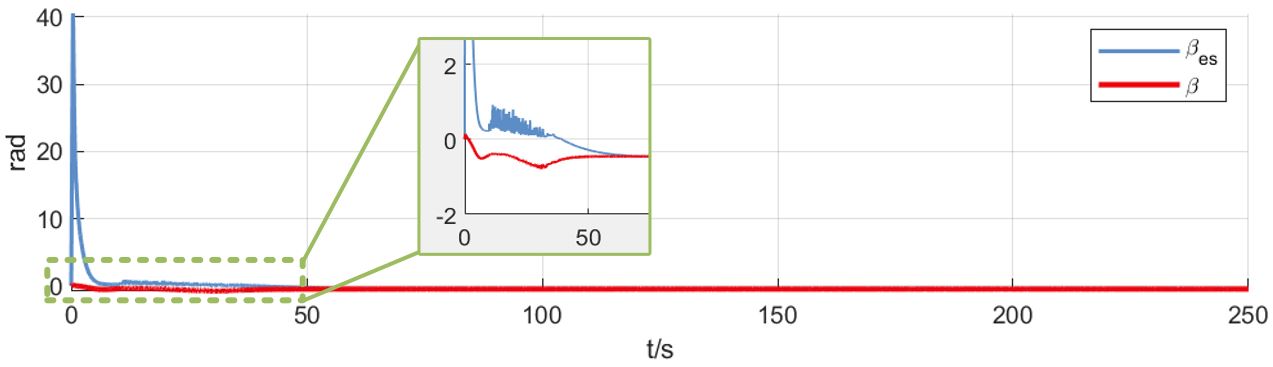} 
}
\subfigure[]{
\includegraphics[width=3in]{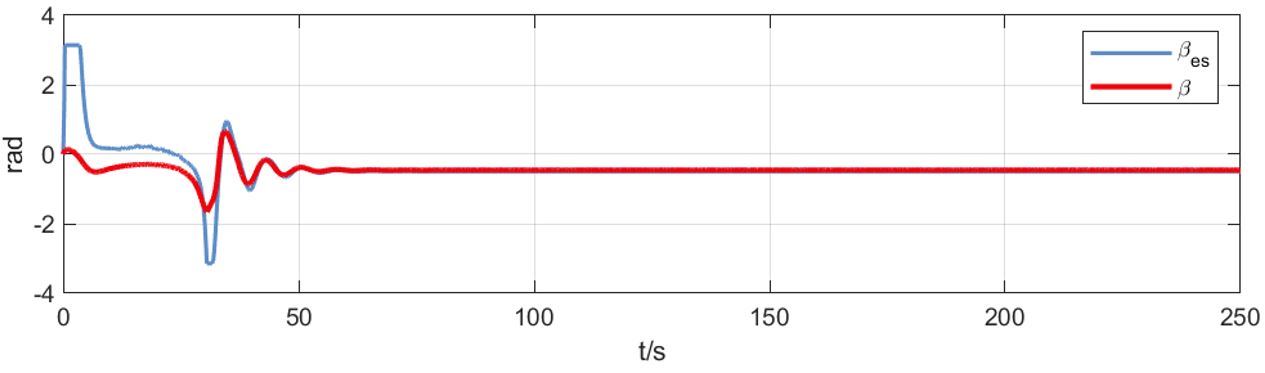} 
}
\subfigure[]{
\includegraphics[width=3in]{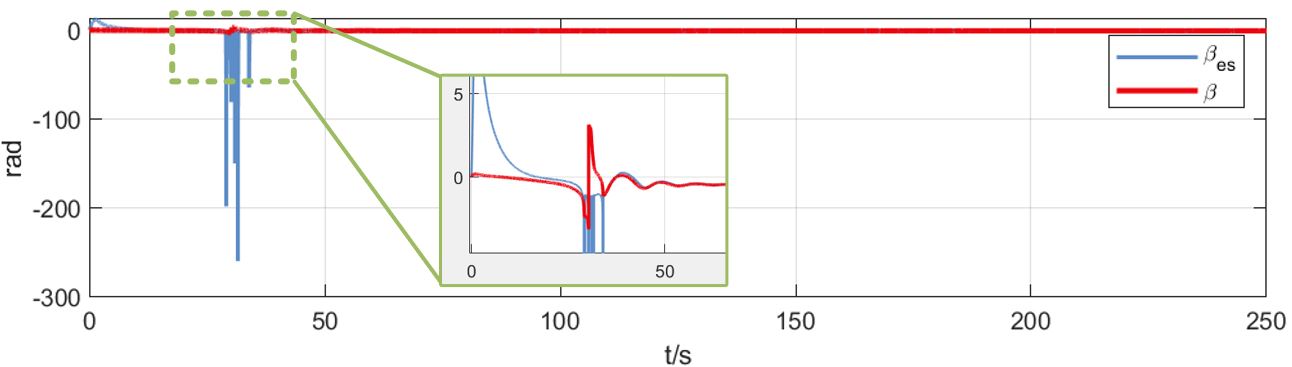}
}
\subfigure[]{
\includegraphics[width=3in]{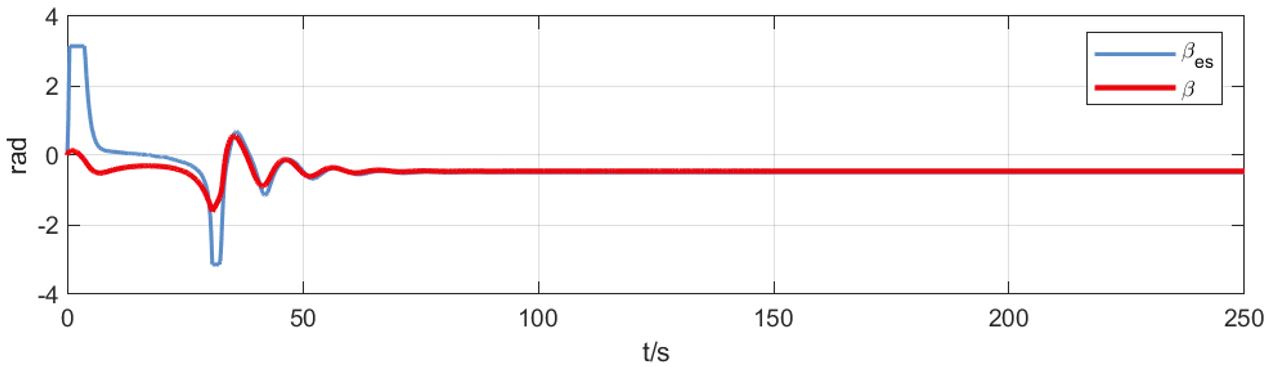}
}
\DeclareGraphicsExtensions.
\caption{The estimation of sideslip angle in case3 $ \sim $ case6}
\label{simulation beta}
\end{figure}
The initial conditions are set as follows: $(x,y)=(0,0)$, $w(0)=0$ and $y_e=10$ with a desired speed of 0.5 m/s. The trajectories and cross-track errors resulting from various guidance laws are depicted in Figure \ref{simulation traj} and Figure \ref{simulation ye}, respectively.
Figure \ref{simulation traj} illustrates that both the LOS and ALOS guidance laws exhibit an overshoot phenomenon, consistent with observations in prior studies such as \cite{Liu2016} and \cite{Fossen2023}. Over time, the ALOS converges towards the desired path, while the LOS continues to display a certain level of deviation. In contrast to the LOS and ALOS, both ELOS and IELOS guidance law are capable of tracking the desired path accurately without any overshoot. Similar phenomenon are also evident in Figure \ref{simulation ye}, where the cross-track error $y_e$ for both LOS and ALOS exhibit overshoot behavior. Notably, the $case4$ displays an abnormal buffeting from 30 s to 40 s which is more pronounced than in the $case6$. This indicates that the transient path tracking capability of ELOS guidance law under low bandwidth condition is inferior to that of IELOS guidance law.

To quantitatively evaluate the performance of various guidance laws, the Mean Absolute Error (MAE) is utilized for both cross-track and along-track errors, as presented in Table \ref{MAE in simulation}. The equation of MAE is presented as follow:
\begin{equation}
{\left( {\begin{array}{*{20}{c}}
{{x_e},}&{{y_e}}
\end{array}} \right)_{MAE}} = \left( {\begin{array}{*{20}{c}}
{\frac{1}{m}\sum\limits_{i - 1}^m {\left| {{x_e}} \right|,} }&{\frac{1}{m}\sum\limits_{i - 1}^m {\left| {{y_e}} \right|} }
\end{array}} \right)
\end{equation}

Table \ref{MAE in simulation} reveals that the MAE ranking for $y_e$ is as follows: $case 1 > case 2 > case 4 $ $ \approx $ $(case 3/ case 5/ case 6)$ while for $x_e$, the ranking is $case 4 > case 1 > case 2 $ $ \approx $ $(case 3/ case 5/ case 6)$. Overall, the performance of ELOS and IELOS guidance laws are superior than LOS and ALOS guidance laws in terms of $y_e$ and $x_e$ except for the $case4$. The IELOS guidance law consistently exhibits high performance regardless of bandwidth, whereas ELOS shows significant variability in performance causing by the bandwidth.

\begin{figure*}[pos=b!]
    \subfigure[]{
		\centering
		\includegraphics[width=18cm]{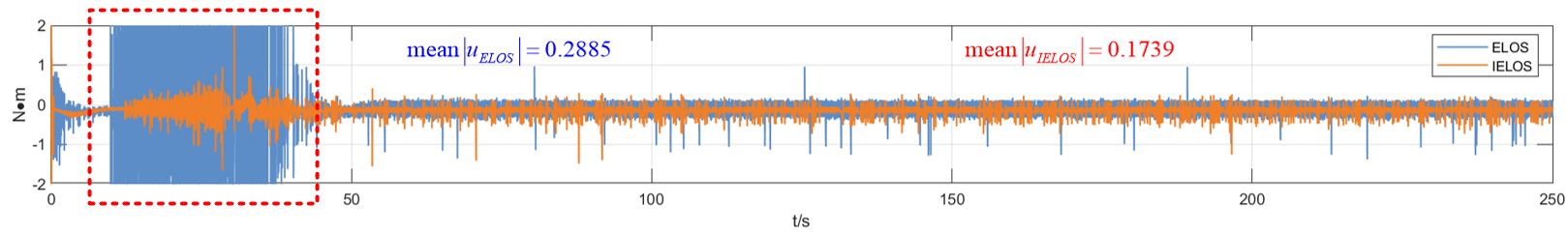} }
    \subfigure[]{
		\centering
		\includegraphics[width=18cm]{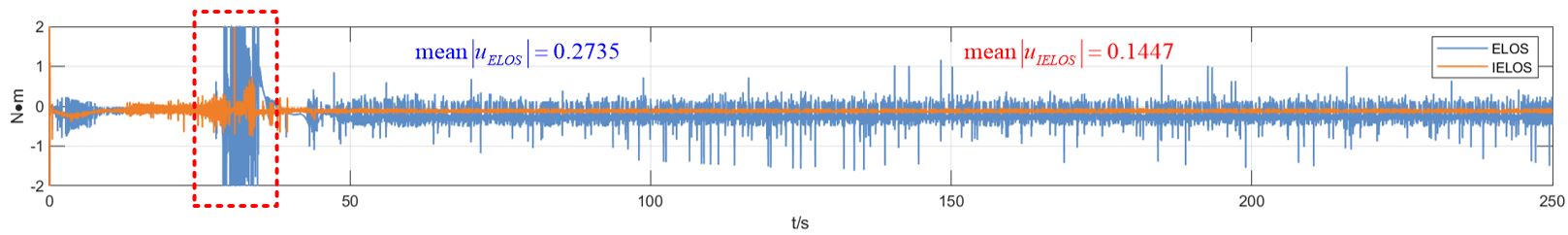} }
\caption{(a) The autopilot torque of case3 and case5. (b) The autopilot torque of case4 and case6}
\label{simulation Tz}
\end{figure*}
Regarding the estimation of slidslip, Figure \ref{simulation beta} illustrates the estimation of sideslip angle in $case3$ $ \sim $ $case6$. In both $case3$ and $case5$, $\beta_{es}$ exhibits buffeting from time 10 s to 30 s, with $case5$ showing noticeable outliers when the true value of $\beta$ undergoes rapid variations (up to 103 degrees per second). In contrast, $case4$ and $case6$ track accurately the $\beta$ when its variation is approximately 74 degrees per second. Obviously, the buffeting phenomenon observed in $case3$ and $case5$ has been eliminated in $case4$ and $case6$ through the utilization of the nonlinear tracking differentiator. Additionally, in $case4$ and $case6$, a noticeable cutoff phenomenon is observed in the estimated sideslip angle $\beta_{es}$ causing by the anti-saturation controller, which helps maintain the $\beta_{es}$ within a reasonable range from $-\pi$ to $\pi$ even when the actual sideslip angle $\beta$ undergoes rapid variations. As a result, the buffeting and outliers observed in ELOS guidance law ($case3$ and $case5$) have been effectively addressed in IELOS guidance law ($case 4$ and $case 6$) through the incorporation of an anti-saturation controller and nonlinear tracking differentiator.

There is a strong correlation between the autopilot torque $u_{\psi}$ and the estimation of sideslip $\beta_{es}$. Consequently, a similar trend is also evident in Figure \ref{simulation Tz}, where IELOS guidance law notably decrease the autopilot torque $u_{\psi}$  and abnormal outliers compared to the ELOS guidance law, regardless of whether bandwidth $k=10$ or $k=2$. Shown as the red dotted box of Figure \ref{simulation Tz} (a) and (b), the buffeting of autopilot torque $u_{\psi}$ has been successfully addressed in transient phase. Additionally, the corporation of nonlinear tracking differentiator efficiently decrease the overall mean torque in steady phase. The overall mean torque of $case3$ to $case6$ has been depicted in Figure \ref{simulation Tz}.

The simulation result indicates that IELOS guidance law can effectively addresses the challenge faced by ESO in accurately tracking sideslip angle with dramatic variation under low observer bandwidth $k$, thereby reducing the buffeting phenomenon observed in ESO. This reduction in buffeting leads to a decrease in the torque generated by the propeller actuators, promoting their long-term durability. Additionally, IELOS outperforms LOS and ALOS, showcasing lower MAE of $x_e$ and $y_e$, and superior overall performance.

\subsection{Pool Experiment}

\begin{table}[width=.9\linewidth,cols=3,pos=h]
\caption{The parameters of various guidance laws in the pool experiment}
\begin{tabular*}{\tblwidth}{@{} LLLL@{} }
    \toprule
          LOS   &  ALOS & ELOS&IELOS\\
    \midrule
         $\Delta=10 $ &  $\Delta=10 $ & $\Delta=10$ &$\Delta=10 $\\
      $\kappa =0.9$& $\kappa =0.9$ & $\kappa =0.9$&$\kappa =0.9$\\
          & $l=0.2$ &$k=0.1$ &$k=0.1$\\
          &         & &$r=30$    \\
          &         & &$h=0.1$   \\
   \bottomrule
\end{tabular*}
\label{parameters in experiment}
\end{table}
\begin{figure}[pos=h!]
    \includegraphics[width=1\linewidth]{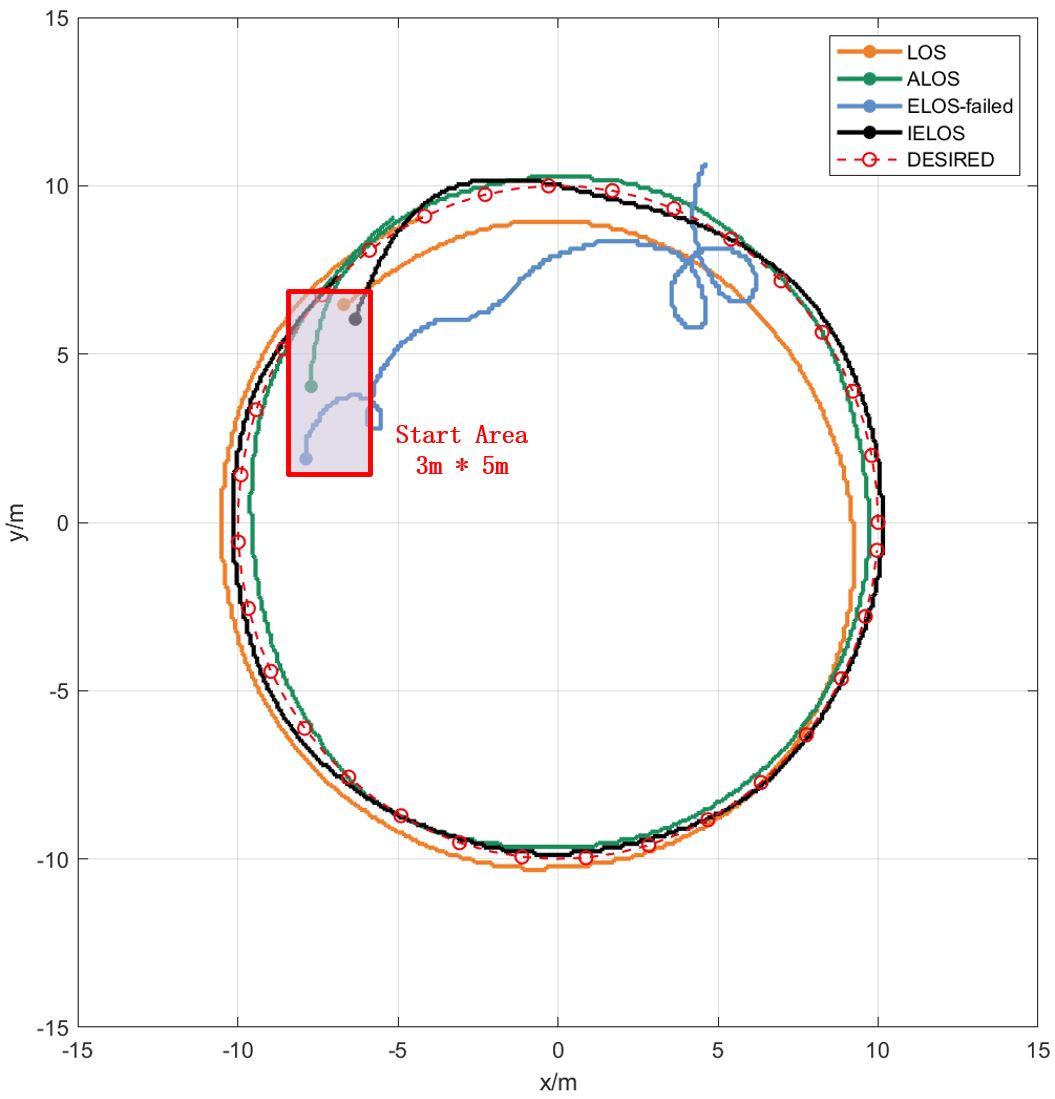}
    \caption{The trajectory by various guidance laws}
    \label{traj_compared}
\end{figure}

\begin{figure*}[pos=b!]
    \centering
    \includegraphics[width=1\linewidth]{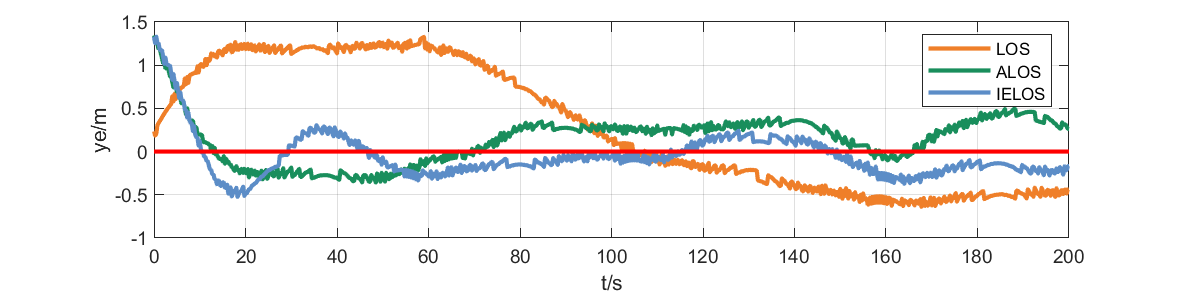}
    \caption{The $y_e$ of LOS, ALOS and IELOS guidance laws}
    \label{expe_ye_compared}
\end{figure*}

\begin{figure}[pos=h!]
    \subfigure[]{
    \includegraphics[width=1.1\linewidth]{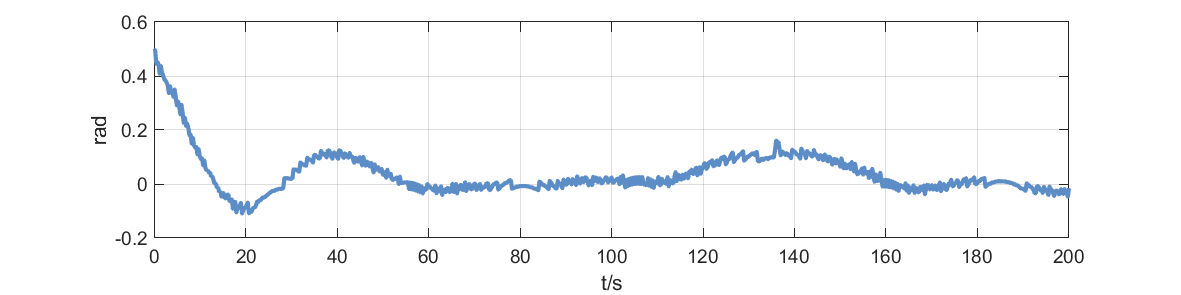}} \subfigure[]{
    \includegraphics[width=1.1\linewidth]{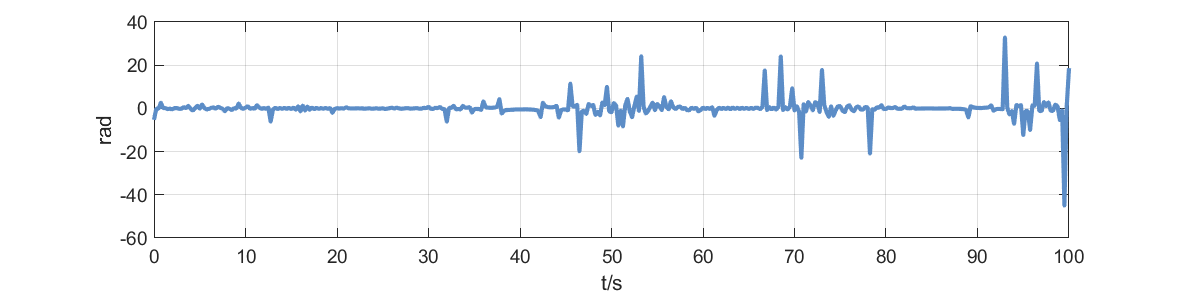}}
    \caption{(a)The estimation of $\beta$ from ELOS, (b)The estimation of $\beta$ from IELOS}
    \label{expe_beta_compared}
\end{figure}

\begin{table}[width=.9\linewidth,cols=3,pos=h!]
\caption{The MAE of $x_e$ and $y_e$ in the pool experiment}
\begin{tabular*}{\tblwidth}{@{} LLL@{} }
    \toprule
             &  $x_e$ (unit: m) & $y_e$ (unit: m)\\
    \midrule
         LOS &  $0.9439$ & $0.7395$\\
         ALOS & $0.0392$ & $0.2641$\\
         IELOS & $0.0309$ & $0.2001$\\

   \bottomrule
\end{tabular*}
\label{MAE in experiment}
\end{table}

\begin{figure}[pos=h!]
\subfigure[]{
    \centering
    \includegraphics[width=1.1\linewidth]{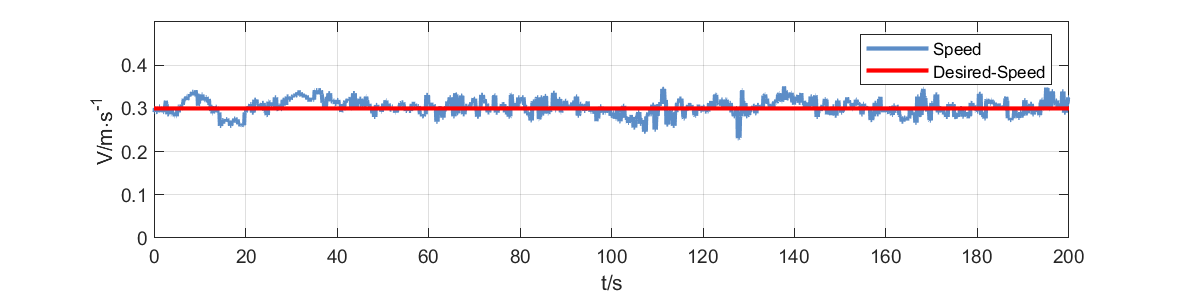}}
\subfigure[]{
    \centering
    \includegraphics[width=1.1\linewidth]{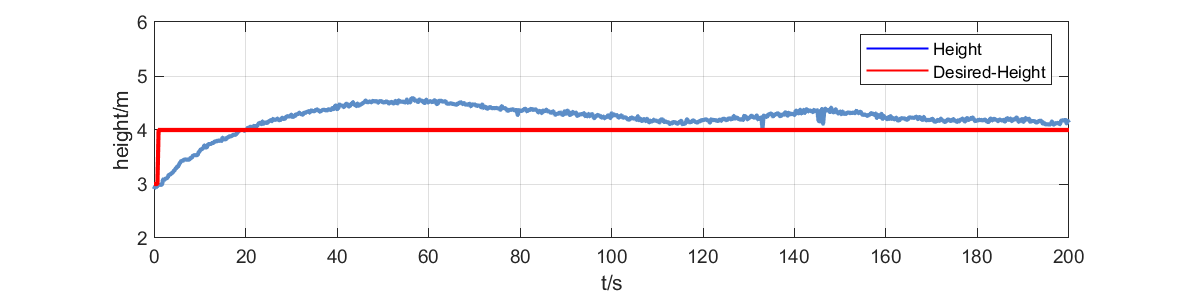}}
    \caption{The speed and height data in pool experiments of IELOS guidance law}
    \label{expe_Height_of_IELOS}
\end{figure}

\begin{figure}[pos=h!]
    \centering
    \includegraphics[width=1\linewidth]{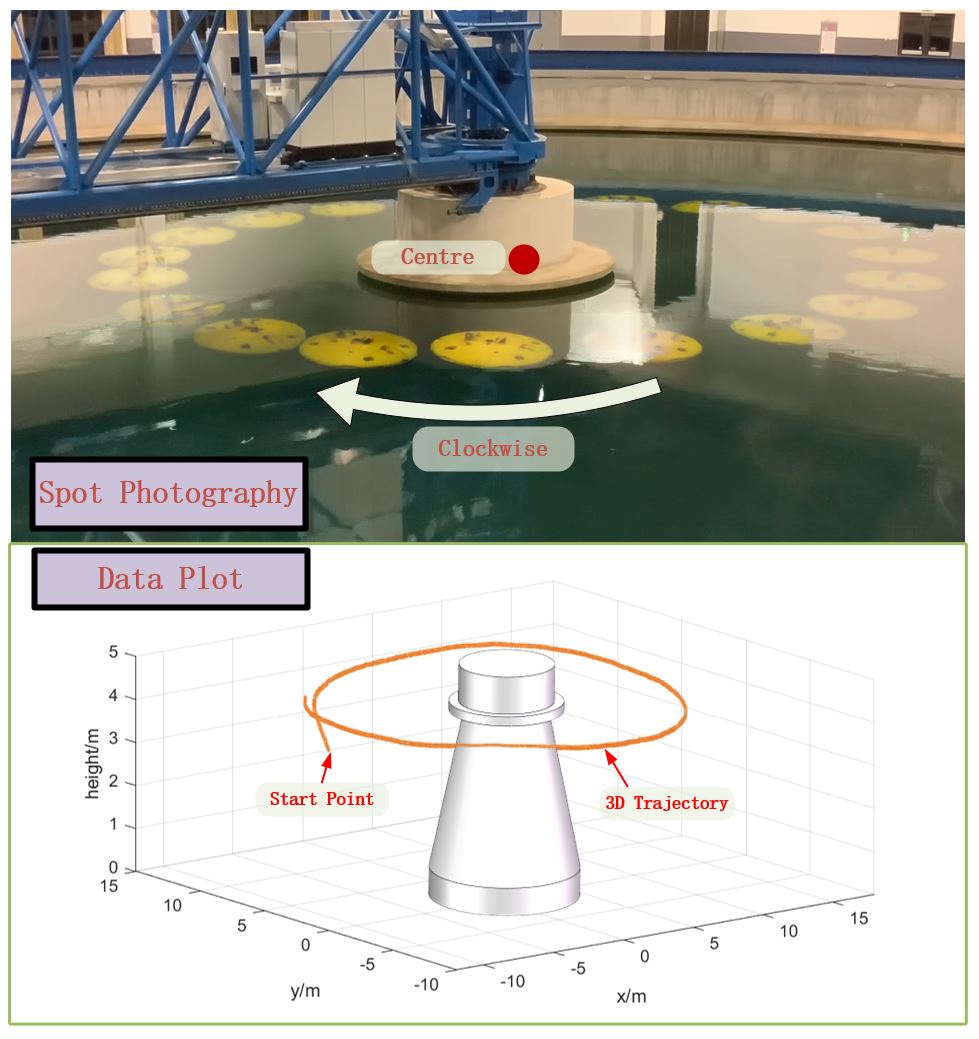}
    \caption{pool experiment}
    \label{pool experiment}
\end{figure}

The pool experiment is conducted in the maneuvering pond at Zhejiang University, China.The pond has a diameter of 45 m and a water depth of 5 m. The primary aim of the experiment is to validate the reliability of IELOS guidance law under low bandwidth conditions and demonstrate its superiority over the LOS and ALOS guidance law.

Similar to the simulation, a circular trajectory with a radius of 10 m is designed as the desired path in the pool experiment. The desired speed is set to 0.3 m/s, and the desired height is 4 m. The parameters for each guidance law are standardized and detailed in Table \ref{parameters in experiment}. The parameters $\Delta$ and $\kappa$ of various guidance laws are set to 10 and 0.9, respectively. Due to the complexity of underwater environment and limitations in sensors accuracy, the bandwidth $k$ is set to 0.1, representing a low value. To simplify experiment's complexity, each guidance law commences from the same general area (3 m $*$ 5 m), although not from the exact same point. The initial condition are established as follows: $w(0)=0$, $\psi=0$, and $z=3$. 

Figure \ref{traj_compared} illustrates that the ELOS guidance law struggles to accurately track the desired circular path and exhibits chaotic trajectory behaviour under the low bandwidth condition, while IELOS and ALOS guidance laws effectively converge to the path. Besides, the LOS guidance law exhibits a evident deviation in steady phase and a overshoot in transient phase, which are consistent with the $case1$ in simulation. Figure \ref{expe_beta_compared} shows that estimated sideslip angle $\beta_{es}$ obtained through IELOS guidance law exhibits smooth and consist behavior within a reasonable range, whereas the corresponding values from ELOS are erratic and fall outside acceptable bounds, ultimately leading to the failure of ELSO guidance law in tracking the desired path. 
Figure \ref{expe_ye_compared} demonstrates that the cross-track error $y_e$ of IELOS guidance law is lower than that of LOS and ALOS. The specifics of $x_e$ and $y_e$ are detailed in Table \ref{MAE in experiment}. This table reveals that the MAE values of $x_e$ and $y_e$ of the IELOS guidance law are the lowest, demonstrating superior performance compared to the LOS and ALOS guidance laws, further confirming IELOS's exceptional tracking accuracy. The 3D trajectory and a third-person perspective view of IELOS are depicted in Figure \ref{pool experiment}. The AUH maintains speed and height control while smoothly tracking the circular path, as shown in Figure \ref{expe_Height_of_IELOS}. Figure \ref{pool experiment} and Figure \ref{expe_Height_of_IELOS} confirms that the AUH excels in coupled motion control. When tracking a curved path, the AUH demonstrates smooth and consistent behavior in heave and speed control, which are essential for 3D curve path following.

In conclusion, the IELOS guidance law outperforms the LOS and ALOS guidance law in the pool experiment, exhibiting lower MAE of $x_e$ and $y_e$. Even under low bandwidth condition, the IELOS guidance law accurately tracks the sideslip angel and converges to the desired circular path, whereas the ELOS guidance law encounters difficulties in estimating the sideslip and results in a chaotic trajectory.

\section{Conclusion and future work}
In general, the performance of the classical reduced-order ESO is strongly correlated with bandwidth: the lager value of bandwidth, the better the ESO performs. However, the bandwidth of ESO is highly dependent on factors such as sampling frequency, sensors accuracy, and uncertain external disturbance, which restricts its value maintained a low level in practical operation. By implementing an anti-saturation controller and nonlinear tracking differentiator, IELOS can successfully tracking the sideslip with significant variation even under low bandwidth conditions, a task unattainable by classical ELOS. Through the simulation and pool experiment, IELOS guidance law showcases its superior by  effectively tracking the sideslip angle without encountering buffeting or outliers, ultimately leading to reduced and consistent autopilot torque. Additionally, IELOS presents lower along-track error $x_e$ and cross-track error $y_e$ compared to LOS and ALOS, further solidifying its superior performance in practical applications.

In future, a field experiment will be undertaken to validate the reliability and anti-interference performance of IELOS guidance law. The experimental platform will utilize the AUH, with IELOS guidance law integrated to facilitate the detection of the wind power platforms, akin to the setup depicted in Figure\ref{pool experiment}. The experiment aims to achieve multi-depth (0m to 100m) and multi-sensor (camera + sonar) 360-degree loop detection capabilities.

\printcredits
\section{Declaration of competing interest}
The authors declare that they have no known competing financial 
interests or personal relationships that could have appeared to influence the work reported in this paper. 

\section{Acknowledgements}
This research was supported by the National Natural Science Foundation of China (Grant No. 52001279). The authors would like to thank Zhikun Wang and other Ocean College people at Zhejiang University for their inspiration and helping experiments.
\bibliographystyle{model1-num-names}
\bibliography{cas-dc}





\end{document}